\documentclass[preprint,aps,psfig,groupedaddress,superscriptaddress,showpacs]{revtex4}
\usepackage{graphicx}
\usepackage{dcolumn}
\usepackage{bm}

\newcommand{\x}{{\bf x}}

\newcommand{\bl}{{\bf l}}
\newcommand{\om}{\omega}

\begin{document}\draft

\title{Statistics of surface gravity wave turbulence in the space and time domains.}

\author{Sergey Nazarenko}
\affiliation{Mathematics Institute, Warwick University, Coventry, CV4 7AL, UK}
\author{Sergei Lukaschuk}
\affiliation{Department of Engineering Hull University, Hull, HU6 7RX, UK}
\altaffiliation[Also at ]{The Institute of Automation and Electrometry, SBRAS,
Novosibirsk} \email{S.Lukaschuk@hull.ac.uk}
\author{Stuart McLelland}
\affiliation{Department of Geography, Hull University, Hull, HU6 7RX, UK}
\author{Petr Denissenko}
\affiliation{School of Engineering, Warwick University, Coventry, CV4 7AL, UK}

\date{\today}

\begin{abstract}

We present experimental results
on simultaneous space-time measurements for the gravity wave turbulence in a large laboratory flume.
We compare these results with predictions of the weak turbulence theory (WTT) based on random
 waves, as well as with predictions based on the coherent singular wave crests.
 We see that both wavenumber and the frequency spectra are not universal and dependent on the
 wave strength, with
 some evidence in favor of  WTT at larger
 wave intensities when the finite flume effects are minimal.
We present further theoretical analysis of the role of the random and coherent waves
in the wave probability density function (PDF) and the structure functions (SFs).
Analyzing our experimental data we found
that the random waves and the coherent
structures/breaks  coexist: the former
show themselves in a quasi-gaussian PDF core
and in the low-order SFs,
and the latter - in the PDF tails and
the high-order SF's.
It appears that the $x$-space signal is more intermittent than the $t$-space
signal, and the $x$-space SFs capture more singular coherent
structures than do the $t$-space SFs.
We outline an approach treating the interactions
of these random and coherent components as a turbulence cycle characterized by
the turbulence fluxes in both the wavenumber and the amplitude spaces.

\end{abstract}

 \maketitle

\section{Introduction}
Understanding statistics of random water surface waves and their mutual
nonlinear interactions mechanisms is important for the wave forecasting,
weather and climate modeling \cite{JanssenBook}. Field observations of the sea
surface, laboratory experiments in wave flumes and numerical simulations are
efficient and complimentary tools for studying such random nonlinear waves and
for testing existing theoretical models. Obvious advantage of the field
observations is that they deal directly with the system we want to know about,
rather than model it in a scaled-down laboratory experiment or in numerical
simulation.
In comparison with field measurements, laboratory experiments and numerical
simulations allow more control over the physical conditions and over the
quantities we measure, especially in the numerical simulations which allows us
to access a much broader range of diagnostics than in experiments.

On the other hand, the laboratory experiments enable observations of much
larger range of wave scales than it is possible in numerical simulations under
the current level of resolution and, therefore, they allow to obtain cleaner
power-law spectra and other scalings. Furthermore, laboratory experiments are
much more realistic than numerics in reproducing the strongly nonlinear events
because most numerical methods are based on weakly nonlinear truncations of the
original fluid equations. Finally, they also have a natural dissipation
mechanism as in open seas, wave breaking, in contrast to an artificial
hyper-viscous dissipation which is usually used in numerics.

In the present paper, we report on new experimental results in a laboratory
flume of dimensions 12m x 6m x 1.5m. In our previous experiments at this
facility \cite{DLN07} we measured time series of the surface elevation at
several fixed locations on the two-dimensional plane using point-like wire
capacitance probes.
This is a standard technique which was also used in smaller experiments
\cite{CW exps}, and which allows one to obtain the wave spectra in the
frequency domain, as well as higher order statistics of the surface heights
from the time series acquired at fixed spatial locations. For weakly nonlinear
waves,
such measurements seem to be sufficient for obtaining information about the
space distributions of the waves (e.g. the wavenumber spectra) via the linear
wave dispersion relation $\omega = \sqrt{gk}$. On the other hand, our previous
experimental results \cite{DLN07} indicated that in the laboratory flumes there
are significant finite-size effects which can be overcome only at rather high
levels of nonlinearity of the wave field. In particular, at high averaged
wave field intensities
we observed a better agreement with the $\omega^{-4}$ prediction of the
Zakharov-Filonenko (ZF) \cite{ZF,ZLF} wave turbulence theory developed for
weakly nonlinear waves with almost random phases. On the other hand, the same
$\omega^{-4}$ spectrum was predicted by Kuznetsov (Ku) \cite{kuznetsov} based
on the assumption that the dominant contribution to the power-law scaling comes
from sharp wave-crests with one-dimensional ridges whose velocity remains nearly
constant while crossing the wire probe. Obviously, the nonlinearity of such
wave-crests is high and one cannot use the linear dispersion relation for
obtaining the space statistics out of the time statistics. This situation
demonstrates that
the frequency spectrum alone does not allow us to distinguish between such
drastically different types of waves: random phased modes and sharp-crested
structures. On the other hand, ZF and Ku theories predict very different shape
of the $k$-spectra, $k^{-2.5}$ (for one dimensional, 1D, spectral density) vs
$k^{-4}$ respectively. Thus, a direct method of measuring $k$-spectra could
allow one to differentiate between the ZF and Ku states.

Besides, even for rather weakly nonlinear on average fields, occasional
strongly nonlinear wave-crests and wave-breaks are known to occur. In spite of
being seldom, these structures are crucial because they provide the main
mechanism for dissipation of the wave energy, and they are related to the
phenomenon of intermittency of wave turbulence. Again, since such seldom events
are strongly nonlinear, one cannot use the linear dispersion relation for
understanding their statistics, and a direct $x$-space measurement is
desirable.

With these motivations in mind, in the present work we have  implemented a new
technique for direct one dimensional measuring of instantaneous surface
profiles
(see details below). Thus, we are able to measure the $k$-spectra directly, as
well as the higher order $x$-space statistics averaged over a discrete set of
instants of time, particularly PDFs and SFs of the height increments in space.
This can be done at different levels of wave forcing, but we have excluded the
range of very weak forcing for which the finite flume size effects were shown
to be
significant.
This was primarily to achieve
better scaling regimes since at low intensities the spectra are very steep and
span over smaller ranges of scales with some peaks often obscuring the power-law
fits \cite{DLN07}. The $x$-space measurements are accompanied by the
capacitance fixed point measurements of the $t$-series for the same
experimental runs, which gives us simultaneous information about the space and
time domain statistics of the water surface elevation.

The main message of the present paper is that the wave turbulence behavior is
 typically
non-universal and reflects presence of two coexisting species: weak incoherent
waves and sparse but  strong sharp wave-crest structures. The incoherent waves
dominate in the spectra and the scalings of the low order structure functions.
These scaling agree with ZF weak turbulence predictions for the spectra at
higher amplitudes. Moreover, we see consistency with ZF scalings for the
low-order SFs at relatively high amplitudes (when the finite size effects
are minimal). Further,  ZF theory appears to agree better with the $t$-domain
than for the $\omega$-domain statistics, which is clearly the finite inertial
range effect (because the $t$- and the $\omega$-objects
are related via the Fourier transform).
 For the higher order SFs, we see behavior
 characteristic of intermittency and presence of singular
 coherent structures. Interestingly, propagating Ku-type 1D wave crests
 are better detected by the $t$-space SFs, and the $x$-space
 SFs capture more singular almost non-propagating wave crests
 schematically shown in Figure \ref{fig:splash}.


 Our paper is organized as follows. In Section two we  describe
 the relevant theories and predictions for the surface wave turbulence.
 In Section three we describe the experimental facility and the measurement
 techniques. In Section four we present the experimental results along with
 their discussion in the context of the theoretical predictions and
 possible interpretations. In Section five we present a summary of our
 findings and an outlook for a future work.

\section{Theoretical background.}

Let us start with a theoretical background including an overview of the
existing theoretical  predictions as well as a further analysis of the
statistical objects relevant to our experiments.

\subsection{Spectra.}

 First, let us define
 the wave energy spectrum in the frequency domain as
\begin{equation}
E_\omega = \int e^{i \omega t'} \langle \eta({\bf x}, t)
\eta({\bf x}, t+t') \rangle \, dt',
\label{Eom}
\end{equation}
and the 1D energy spectrum in the wavenumber domain, respectively, as
\begin{equation}
E_k = \int e^{i k z} \langle \eta({\x} , t)
\eta({\x} + {\bf w} z, t) \rangle \, dz,
\label{Ek}
\end{equation}
where $\eta({\bf x}, t)$ is the surface elevation at time $t$ and location in
the horizontal plane ${\bf x}=(x,y)$. The integration in (\ref{Eom}) is taken
over a time window, and in (\ref{Ek}) - over a piece of straight line in the 2D
plane illuminated by the laser sheet (with ${\bf w}$ being a unit vector along
this line).
  angle brackets mean ensemble
averaging over realizations (equivalent to the time averaging in presence of
ergodicity). For a statistically steady and homogeneous state, $E_\omega$ and
$E_k$ are independent of $t$ and ${\bf x}$. Most of the theories predict a
power-law scaling
\begin{equation}
E_\omega \propto \omega^{-\nu}
\label{nu_power_law}
\end{equation}
and
\begin{equation}
E_k \propto k^{-\mu} \label{mu_power_law}
\end{equation}
where the indices $\nu$ and  $\mu$  depend on a particular theory.

\subsection{Statistics of the field increments.}

The spectra considered in the previous section correspond to the second-order
correlators.  Different types of coherent and incoherent structures may lead to
the same spectra. One, therefore,  must consider higher-order correlators
to see an unequivocal signature of a particular kind of coherent structures or
incoherent random phased field.

To study the higher-order statistics in our previous paper \cite{DLN07} we
considered PDFs of the wave crest heights, as well as PDFs of a band-pass
filtered field (the theory for the later was developed in \cite{clnp}). Here,
we will also consider other (very popular in turbulence theory) objects: space and
time elevation field increments of different orders which are defined as
follows,
\begin{equation}
\delta^{(1)}_{\l}  = \eta (\x + \bl) - \eta (\x), \label{delta1_l}
\end{equation}
\begin{equation}
\delta^{(2)}_{\l}  = \eta (\x + \bl) - 2\eta (\x) + \eta (\x -\bl),
\label{delta2_l}
\end{equation}
etc. (here all $\eta$'s are taken at the same $t$), and
\begin{equation}
\delta^{(1)}_\tau  = \eta (t+\tau) - \eta (t),
\label{delta1_tau}
\end{equation}
\begin{equation}
\delta^{(2)}_\tau  = \eta (t+\tau) - 2\eta (t) + \eta (t-\tau) ,
\label{delta2_tau}
\end{equation}
etc. (here all $\eta$'s are taken at the same $x$).

\subsubsection{Probability density functions.}

PDFs of the above increments $P_{x} (\sigma)$ and  $P_t (\sigma)$ are defined
in the usual way as a probability of a particular increment to be in the range
from $\sigma$ to $\sigma+ d \sigma$ divided by $d \sigma$, or in the symbolic
form
\begin{equation}
P^{(j)}_{x} (\sigma) = \langle \delta (\sigma - \delta^{(j)}_l  ) \rangle,
\label{pdfx}
\end{equation}
and
\begin{equation}
P^{(j)}_{t} (\sigma) = \langle \delta (\sigma - \delta^{(j)}_\tau  ) \rangle
\label{pdft}
\end{equation}
respectively, where $j =1,2, \dots$. For random phased fields, these PDFs are
Gaussian, and presence of sparse coherent structures can be detected by the
deviations from Gaussianity at the PDF tails. In particular, fatter than
Gaussian tails indicate an enhanced probability of strong bursts in the signal
which is called intermittency.

\subsubsection{Structure functions.}

Let us now introduce the moments of the height increments, which are called the
{\em structure functions},
\begin{equation}
S^{(j)}_l(p) = \langle (\delta^{(j)}_l )^p \rangle = \int \sigma^p  P^{(j)}_{x} (\sigma) \,
d\sigma,
\label{SFx}
\end{equation}
and
\begin{equation}
S^{(j)}_\tau(p) = \langle (\delta^{(j)}_\tau )^p \rangle = \int \sigma^p  P^{(j)}_{t} (\sigma) \,
d\sigma.
\label{SFt}
\end{equation}
Often in turbulence, the structure functions asymptotically tend to scaling laws,
\begin{equation}
S^{(j)}_l(p) \sim l^{\xi(p)}
\label{SFx_scale}
\end{equation}
in the limit $l \to 0$, and
\begin{equation}
S^{(j)}_\tau(p) \sim \tau^{\zeta(p)}
\label{SFt_scale}
\end{equation}
in the limit $\tau \to 0$ respectively. Functions $\xi(p)$ and $\zeta(p)$ are called
the SF scaling exponents, and they contain the most important information about the
turbulent field coherent and incoherent components and, correspondingly, about the
turbulence intermittency.

\subsection{Scalings generated by random phased waves.}

\subsubsection{Spectra.}

Weak turbulence theory (WTT) considers weakly nonlinear random-phased waves in
an infinite box limit. For the wave spectrum, these assumptions lead to the
so-called  Hasselmann equation \cite{hasselmann}. This equation is quite
lengthy and for our purposes it suffices to say that the ZF energy spectrum
\begin{equation}
E_\omega \propto \omega^{-4}
\label{zf_om}
\end{equation}
 is an exact solution
of Hasselmann equation which describes a steady state with energy cascading
through an inertial range of scales from large scales, where it is produced, to
the small scales where it is dissipated by wavebreaking. In the $k$-domain, the
one-dimensional ZF energy spectrum is
\begin{equation}
E_k \propto k^{-5/2}.
\label{zf_k}
\end{equation}

 It is  important that in deriving WTT, the limit of an
infinite box is taken before the limit of small nonlinearity. This means that
in a however large but finite box, the wave intensity should be strong enough
so that the nonlinear resonance broadening is much greater than the spacing of
the $k$-grid (corresponding to Fourier modes in a finite rectangular box). As
estimated in \cite{sandpile}, this implies a condition on the minimal angle of
the surface elevation
$\gamma > 1/(kL)^{1/4},$
where $L$ is the size of the basin, which is quite a
 severe restriction.
If this condition is not satisfied the number of exact and quasi four-wave
resonances will be drastically depleted
\cite{kartashova1,kartashova2,sandpile}. This can lead to a significant
slowdown of the energy cascade from long to short waves and, therefore, a
steeper energy spectrum. A theory of discrete wave turbulence developed in
\cite{sandpile} for very low levels of forcing predicts $E_\omega \propto
\omega^{-6}$, which is confirmed in experiments with very weak forcing
\cite{DLN07}. On the other hand, for such weak forcing the scaling interval is
rather short and not very well formed (it contains some peaks). Thus, in the
present paper we will deal with stronger wave fields for which the spectra are
shallower ($\nu \lesssim 5.5$), even though it is still steeper than ZF due to
the finite-size effects.

\subsubsection{PDF and the structure functions.}

Now let us consider a wave field made out of modes with random phases and the
energy spectrum $E_k \sim k^{-\mu}$. As we mentioned, for random phased fields,
the PDFs of the height increments
 are Gaussian. For Gaussian statistics
we immediately have
\begin{equation}
S^{(j)}_l(p) \sim l^{p(\mu -1)/2}
\label{sf_rf}
\end{equation}
 if $\mu < 2 j + 1$, otherwise
$ S^{(j)}_l(p) \sim l^{pj}
$
because the field is $j$ times differentiable.

Similarly, in the time domain we have for the random-phased field with
the energy spectrum $E_\om \sim \om^{-\nu}$:
\begin{equation}
S^{(j)}_\tau(p) \sim \tau^{p(\nu -1)/2}
\label{sf_rf_om}
\end{equation}
 if $\nu < 2 j + 1$, otherwise
$ S^{(j)}_\tau(p) \sim \tau^{pj}
$.

\subsection{Scalings generated by singular coherent structures.}

\subsubsection{Spectra.}

Sharp wave crests are quite common for gravity waves of sufficiently large amplitudes.
  The most common type of crests discussed in the literature
  looks like a break in the surface slope.
 A prototype for such structures is a
sharp-crested stationary Stokes wave solution with the crest angle of
$120^\circ$. Following Kadomtsev \cite{kadomtsev}, such sharp crested waves are
usually associated with the Phillips (Ph) spectrum. Indeed, assume
 that  there are  discontinuities  occurring at  isolated points.
This leads to the following   one-dimensional
energy spectrum in wavenumber space
$E_k \propto k^{-3}.$
Second, assuming that transition from the  $k$-space to the $\omega$-space
should be done according to the linear wave relation $\omega = \sqrt{gk}$, we
arrive at the Ph spectrum \cite{phillips},
\begin{equation}
E_\omega = g^{2} \omega^{-5}.
\label{ph}
\end{equation}
An alternative way to derive the Ph spectrum, the way it was originally done by
Phillips \cite{phillips},  is to assume that the gravity constant $g$ is the
only relevant  dimensional physical quantity. This argument is equivalent to
saying that the linear term is of the same order as the nonlinear one in the
water surface equations in the Fourier space.

Kuznetsov  \cite{kuznetsov} questioned this picture and argued that (i) slope
breaks occur on one-dimensional lines/ridges rather than on zero-dimensional
point/peaks, and (ii) that the wave-crest is propagating with preserved shape,
i.e. $\omega \propto k$ should be used instead of the linear wave relation
$\omega = \sqrt{gk}$. This assumptions give $E_\omega \propto \omega^{-4}$,
i.e. formally the same scaling as ZF, even though the physics behind it is
completely different. Finally, it was proposed in \cite{cnn} that wave crest
ridges may have non-integer fractal dimension $D$ somewhere in the range
$0<D<2$. This leads to the following   one-dimensional energy spectrum in the
$k$-space,
\begin{equation}
E_k \propto k^{-3-D}.
\label{frac_k}
\end{equation}
Assuming, following Kuznetsov,  $\omega \propto k$, we
have in this case
\begin{equation}
E_\omega \propto \omega^{D-6}.
\label{fractal}
\end{equation}

\subsubsection{PDF and the structure functions.}

\begin{figure}
\includegraphics[width=16cm]{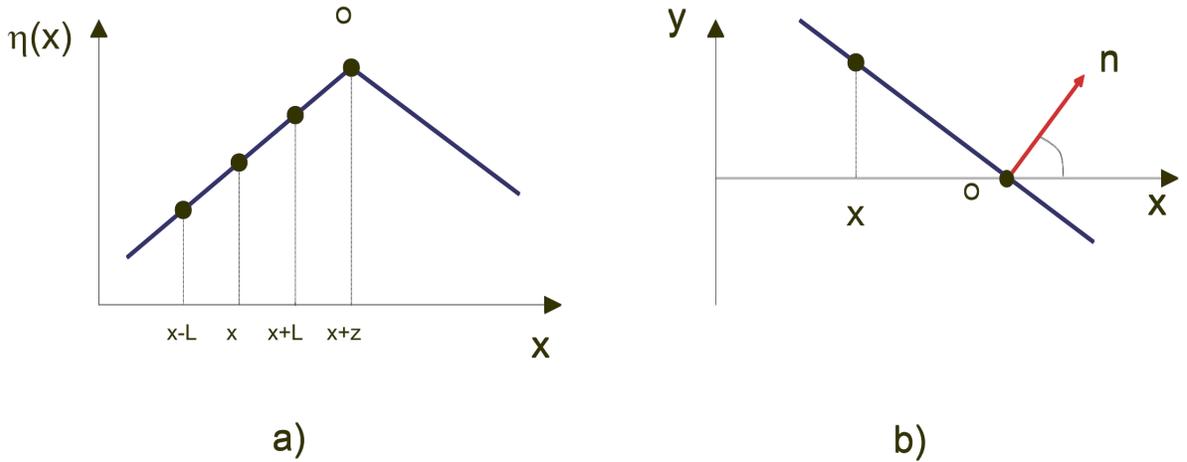}
\caption{\label{fig:sf} Schematic of the ridge (coherent structure) cross
section used in PDF calculations: a) the Ku-type ridge in the plane
$(x,\eta(x))$, where $x$ is directed along the illuminated line on the
air-water interface and is parallel to the vector $\bl$; b) the same ridge in
the $(x,y)$-plane.}
\end{figure}

Let us ask what shape of PDF would be produced by structures  of Kuznetsov (Ku)
 type with unit slopes on the
both sides and $D=1$.
Let us assume that the position and
orientation of such ridges are random in the 2D plane. In this case for $j \ge
2$, a space height increment will have non-zero values only if it has its
argument points on both sides of the coherent structure. Let us restrict
ourselves to the case $j=2$; then
\begin{eqnarray}
\delta^{(2)}_{\l}  &= 2 \cos \theta \;  (l -z), \;\; \hbox{for} \;\; |z| < l, \nonumber \\
\delta^{(2)}_{\l}  &= 0,  \;\;\;\;\;\;\;\;\;\;\;\;\;\;\;\;\;\; \hbox{for} \;\; |z| > l,
\label{delta1_lst}
\end{eqnarray}
where $l = |\bl|$, vector $\bl$ is parallel to the axis $x$, $\theta$ is the
angle between the normal vector to the coherent structure's ridge and vector
$\bl$,
 and $z$ is the distance from the increment middle point and the intersection between the ridge and
 the line connecting the increment points, see Fig. \ref{fig:sf}.

For the PDF we have
\begin{eqnarray}
P^{(2)}_{x} (\sigma) =0, \;\; \hbox {for} \;\;  |\sigma| > 2 l \label{pdfx1a}
\end{eqnarray}
 (the maximal value of the increment is limited
by the strength of the singularity), and
\begin{eqnarray}
P^{(2)}_{x} (\sigma)   = \langle \delta (\sigma - \delta^{(2)}_l) \rangle =
\;\;\;\;\;\;\;\;\;\;\;\;\;\;\;\;\;
\;\;\;\;\;\;\;\;
\;\;\;\;\;\;\;\; \nonumber \\
{1 \over \pi} \int_{-\pi/2}^{\pi/2} d \theta \; {1 \over 2 l} \int_{-l}^l dz \;\;\;\;
\delta (\sigma  -
2 \cos \theta \;  (l -z)) =
\nonumber \\
{1 \over 2\pi l} \int_{0}^{\arccos {|\sigma| \over 2l}} {d \theta \over
 \cos \theta} =
 {1 \over 2\pi l} \ln \left[ {2l \over |\sigma|} + \sqrt{\left[ {2l \over |\sigma|}
  \right]^2 -1} \,\right]
\label{pdfx1b}
\end{eqnarray}
for $|\sigma| \le 2 l$.

In the above, we considered a simplified configuration of ridges
such that on average only one ridge with slope equal to $\pm 1$ crosses a unit
area. This result can be easily extended to the structures with a distribution
of the slope values and with an arbitrary density in 2D space. This would move
the PDF cutoff to $|\sigma| = 2ls$, where $s$ is the maximal allowed slope of
the coherent structures. (Such a PDF cutoff feature was discussed in
\cite{clnp}; see also discussion below in the next subsection).
Asymptotic behavior for such general PDF for $|\sigma| \ll 2 l$ is
\begin{eqnarray}
P^{(2)}_{x} (\sigma) \sim {A \over l} \, \left[ \ln \left( {l \over |\sigma|} \right)
+B \right],
\label{pdfx1as}
\end{eqnarray}
where $A$ and $B$ are dimensionless constants which depend on the
strength distribution of the singular ridges and their spatial density.

\begin{figure}
\includegraphics[width=\linewidth]{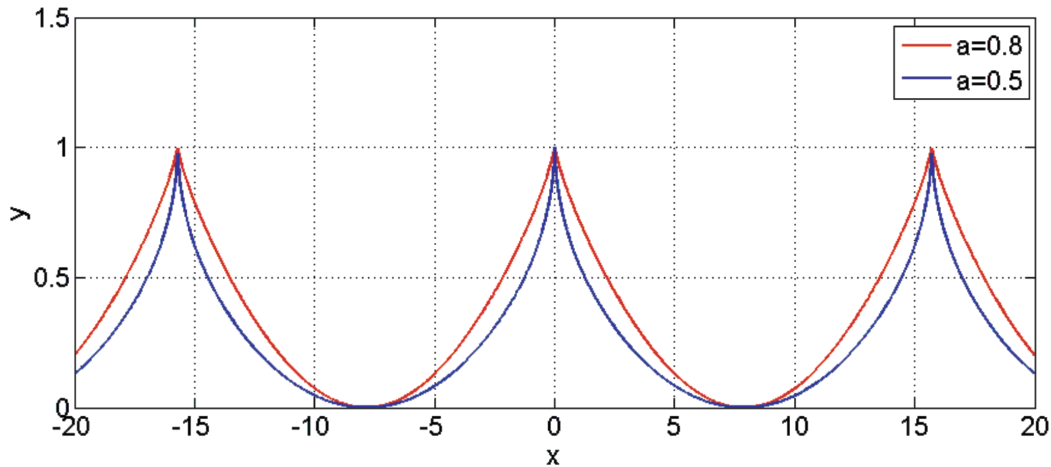}
\caption{\label{fig:splash} A wave profile with singular structures of type
$\eta(x) = \eta_0 - C_a \, |x|^a$ with $0<a<1$. (Plot shows
$y=1-|\sin(0.2x)|^a$.)}
\end{figure}

Let us now consider a somewhat more general class of singular coherent
structures whose cross-section near the singularity is given by formula
\begin{equation}
\eta(x) = \eta_0 - C_a \, |x|^a,
\label{spike}
\end{equation}
with a singularity degree  constant $a$ such that $0<a\le 1$, and constants
$\eta_0$ and $C_a$ describing a reference surface elevation and the coherent
structure amplitude respectively, see Fig.\ref{fig:splash}. For the simplified
structures considered above $a=1$ and $C_a\approx1$ which was implied by both
the Ph and Ku models.
We will see below that structures with the additional parameter $a<1$ also seem
to be relevant to the wave turbulence in our experiments. For generalization we
will assume that the ridges of such crests have a fractal dimension $0\leq D<2$
(e.g. 0 for Ph and 1 for Ku).

Contribution  to the PDF of the structures with $a<1$ can be considered as
above, and it is easy to see that the corresponding integral of type as in the
Eq.\ref{pdfx1b} is convergent. This means that the structures with $a<1$
contribute to the PDF tails "locally", i.e. structures with amplitudes $C_a$
form a "bump" on the tail near $\sigma = C_a l^a$. The shape of such a bump is
not universal and it depends on the distribution of the crests over $C_a$.

On the other hand, an effective way to detect the structures is to consider the
SFs and their scalings in $l$ (note that the PDFs are usually measured at fixed
$l$'s and, therefore, the corresponding scalings in $l$ are hard to obtain,
e.g. due to insufficient statistical data). We will be interested in the limit
$l \to 0$  of the structure functions $S^{(j)}_l(p)$. Suppose that there are
$N$ coherent structures per unit area. The probability for having  points of
the structure function on the two different sides of the coherent structure
ridge is $N  l^{2-D}$. The probability to have all the point on the same side
of the ridge is $1-  l^{2-D} \approx 1$ (for $l \to 0$). If all the points are
on the same side of the ridge then, assuming that away from the ridge the field
is $j$ times differentiable,
$\delta^{(j)}_{\l}  \sim l^j$, whereas if the points are on the different sides
of the ridge $\delta^{(j)}_{\l}  \sim l^a$, so
\begin{equation}
S^{(j)}_l(p) \sim l^{pj} + N  l^{2-D + ap}.
\label{sf_struc}
\end{equation}
Note that in the limit $l \to 0$ out of the two terms on the RHS the one with
the smallest power will be dominant. Thus, the structures of Ph or Ku type,
i.e. with $a=0$ and $a=1$, will not bee seen in SFs for the first-order
increments and we would have to consider $j \ge 2$. However, one should keep in
mind that the finite range of excited scales makes determination of the
scalings less precise for higher orders $j$ because of the higher number of SF
points to be placed in this finite range. Therefore, it is better to consider
the lowest $j$ that could allow to extract the scalings induced by the coherent
structures ($j=2$ in case of the Ph and Ku).

Now suppose that the wave field is bi-fractal and
consists of two components: random phased modes
and singular coherent structures. Avoiding the choices of $j$ for which the field is
$j$ times differentiable, we have in this case
\begin{equation}
S^{(j)}_l(p) \sim l^{p(\mu -1)/2} +  l^{2-D + ap}.
\label{sf_sum}
\end{equation}
If $a<(\mu -1)/2$ we expect to see the scaling associated with the incoherent  random
phased component at low $p$'s (first term on the RHS) and the singular coherent
structure scaling at high $p$'s (second term on the RHS).

Similarly, one can consider the SFs of the time increments. Assuming following
Kuznetsov that the coherent structures could be thought as passing the wire
probes with constant velocity (due to shortness of the time needed for the
singular ridge to pass the probe), we should obtain the time-domain scalings to
be identical to the space-domain scalings obtained above, i.e. $ S^{(j)}_\tau
(p) \sim \tau^{2-D + ap}$.

In the case when incoherent waves and singular coherent
structures are present simultaneously,  we have
\begin{equation}
S^{(j)}_\tau (p) \sim \tau^{p(\nu -1)/2} + \tau^{2-D + ap}.
\label{sf_sum_time}
\end{equation}
As before, it is understood here that the order $j$ is chosen
in such a way that the field associated with the incoherent
wave component is not $j$ times differentiable in time.
For example, for spectra  with $3 < \nu < 5$ (e.g. for ZF spectrum)
one should use $j \ge 2$, and for
$5 < \nu < 7$ one
should use $j \ge 3$, etc.

\subsection{Turbulence cycle and fluxes in the wavenumber-amplitude space.}

In the previous sections we used the Fourier space for the spectra whereas the
higher order statistics was described in terms of the $x$ and $t$-domain
increments. In this subsection, we will outline how one can put turbulence
containing of incoherent waves and coherent structures onto the same "map".
Namely, we will be interested in a turbulence cycle where the structures arise
from the turbulent cascade of incoherent waves and, in turn, incoherent waves
arise during breaking of the structures. The key element of this picture is
combining fluxes over wavenumbers (associated with the Kolmogorov-Zakharov
cascade states) and over the wave amplitudes (considered in \cite{clnp} and
linked to intermittency).

Let us summarize the findings of  \cite{clnp}, where the WTT formalism was
extended to PDF of Fourier intensities $J_k =|a_k|^2$  which is defined as
\begin{equation}
\mathcal{P}_k(J) = \langle \delta(J-|a_k|^2) \rangle.
\label{kpdf}
\end{equation}
Under the usual WTT assumptions (weak nonlinearity, random phases and
amplitudes of the Fourier modes), the following equation for such a PDF was
derived,
\begin{equation}
\dot \mathcal{P} +\partial_{J}F=0, \label{peqn}
\end{equation}
 where
 \begin{equation}
  F=-J(\beta \mathcal{P} +\alpha \partial_{J} \mathcal{P})
  \label{Feqn}
\end{equation}
 is a probability flux in
the $J$-space, and
\begin{eqnarray}
\alpha_k = 4 \pi \int |W({\bf k}, {\bf k}_1, {\bf k}_2, {\bf k}_3)|^2
\delta({\bf k} + {\bf k}_1 - {\bf k}_2 - {\bf k}_3)
\delta(\omega_k + \omega_{k_1} - \omega_{k_2} - \omega_{k_2}) n_{k_1} n_{k_2} n_{k_3} \, d{\bf k_1} d{\bf k_2} d{\bf k_3},\nonumber
\\
\beta_k =
8 \pi \int  |W({\bf k}, {\bf k}_1, {\bf k}_2, {\bf k}_3)|^2
\delta({\bf k} + {\bf k}_1 - {\bf k}_2 - {\bf k}_3)
\delta(\omega_k + \omega_{k_1} - \omega_{k_2} - \omega_{k_2}) \hspace{3.5cm} \nonumber
\\
 \Big[ n_{k_1} (n_{k_2} + n_{k_3}) - n_{k_2} n_{k_3}\Big]  \, d{\bf k_1} d{\bf k_2} d{\bf k_3},  \hspace{2cm} \nonumber
\end{eqnarray}
where $n_k = \langle J \rangle$ and $W({\bf k}, {\bf k}_1, {\bf k}_2, {\bf k}_3)$ is the nonlinearity
coefficient which for the case of the surface gravity waves can be found in
\cite{krasitskii}.
It was shown that there are solutions for such wave PDFs that have power-law
tails and which, therefore, correspond to the states with turbulent
intermittency. These solutions correspond to a constant $J$-flux of
probability, $ F= \hbox{const}$. At the tail of the PDF, $J \gg \langle J
\rangle = n_k$, the solution can be represented as series in $n_k /J$,
\begin{equation}
\mathcal{P}_k(J) = - F/( J \beta)-\alpha F/(\beta J)^2+\cdots.
 \label{1/J}
\end{equation}

It was speculated that such a flux in the amplitude space can be
physically generated by the wave breaking events.
On the other hand, we know that the ZF state (and, in general, KZ spectra
in other applications) corresponds to the energy flux through wavenumbers
$k$. Below, we will consider a combined flux which has both $k$ and $J$ components and,
thereby, clarify the picture of the wave turbulence cycle which involves both
random waves and coherent structures which can get transformed into each other.

\begin{figure}
\includegraphics[width=10cm]{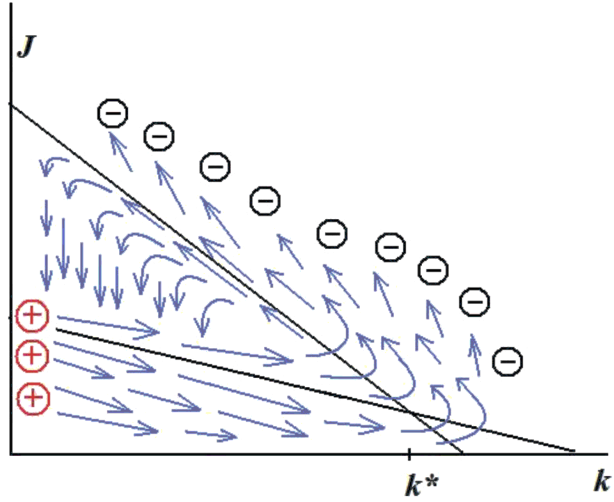}
\caption{\label{WTcycle} Turbulent probability fluxes in the  $(k,J)$-plane.}
\end{figure}

First note that the situation is quite subtle because the intermittent
solution corresponds to the negative $J$-flux, i.e. from large to small
amplitudes, whereas naively one would expect the opposite direction
based on the picture that the wave breaking occurs and dissipates turbulence
when amplitudes become large.
To resolve this "paradox", one should remember that when amplitudes become
large the all $k$-modes become correlated (i.e. we observe occurrence of the
coherent structures). Namely, these modes are concentrated at the following lines
in the $(k,J)$-plane,
$$
J_{Ph} \sim g^{1/2} k^{-9/2}, $$
see Figure \ref{WTcycle}. Note that the subscript Ph here stands for "Phillips"
to emphasize that it corresponds to the Phillips scaling where the
linear and nonlinear terms are of the same order.
Let us consider the fluxes on the $(k,J)$-plane.
Let us force turbulence by generating weak waves at low $k$'s, - region
marked by $\bigoplus$ in Figure \ref{WTcycle}.
The energy cascade will proceed from the forcing region to higher $k$
predominantly along the curve $J(k) =\langle J_k \rangle = n_k$.
For example, for  the ZF state this curve is
$$
J(k) = n_{ZF}(k)  \sim \epsilon^{1/3} k^{-4}. $$

Around some scale $k_*$, where the Ph and ZF curves intersect, the WTT
description breaks down because the nonlinear term becomes of the same order as
the linear one. At this point, the phases get
 correlated, which arises in the form of coherent structures in the
 $x$-space. Such coherent structures are made of a broad range of
 Fourier modes which are correlated and for each of whom the
 linear and nonlinear terms are in balance. Indeed, if the linear term
 for some $k$ was greater than the nonlinear term, then this wave
 would quickly de-correlate from the rest of the modes. If, on the other hand,
 the nonlinear term gets larger than the linear one at some $k$, then
 the inertial forces on a fluid element would get larger than the gravity
 force, and this fluid element would separate from the surface and
 exit the coherent structure.
 However, such sea spray and foam do form occasionally via wave breaking
 which provides the main mechanism of the wave energy dissipation.
Thus, on the $(k,J)$-plane, the flux turns at $k_*$ and goes back to
lower $k$'s along the Ph curve, with some energy lost to the regions
above the Ph curve via wave breaking.
Occasionally, the coherent structures lose their coherence due to the energy
loss to the sea spray and foam and corresponding reduction in nonlinearity.
Additional mechanisms that can promote de-correlation of  coherent structures and due
to their mutual interactions and due to interactions with the incoherent
component. On the $(k,J)$-plane, this corresponds to turning of the flux
down below the Ph curve and toward the ZF curve.
This closes the cycle of the wave turbulence, in which the energy
cascade of the random phased waves leads to creation of coherent structures, which
in turn, break down with their energy partially dissipated in whitecapping
and partially returned into the incoherent random phased component.
The exact partition of the energy dissipated vs the  energy returned into
the random waves is not known, but it is natural to think that these
parts are of the same order of magnitude.

The last part of the wave turbulence cycle is crucial for understanding intermittency.
Indeed, this part corresponds to a flux in the opposite $J$-direction, which, as we
mentioned, corresponds to the power-law tails of the PDF of the Fourier modes.

Note that the wave turbulence cycle picture similar to the one described above
was previously suggested in \cite{dlnz}  in the context of the inverse cascade
in optical wave turbulence in NLS model with focusing nonlinearity (see also a
detailed description in the concluding section of \cite{nnb} and in a recent
paper \cite{N&Z}). In this case, the inverse waveaction cascade  proceeds
within the incoherent weakly nonlinear wave component until it reaches some
(low) wavenumbers where the nonlinearity ceases to be small and the
modulational instability sets in. The modulational instability evolves into
wave collapses which are strongly nonlinear singular events shrinking to a very
small spatial size in finite time. The collapse dissipates part of the
waveaction supplied to it via the inverse cascade, whereas the remaining
waveaction returns back to the system of random waves because the collapse
spike has a significant high-$k$ component which becomes incoherent when after
the collapse burn-out.

The qualitative picture of the wave turbulence cycle outlined above can yield
to some important qualitative predictions.
From the definition of the energy flux in the $k$ space, $\epsilon_k$,
and by taking the first moment of equation (\ref{peqn}), he have
\begin{equation}
\dot E_k = -\partial_k \epsilon_k = 2 \pi k \, \omega_k \int_0^{J_{Ph}} J \partial_J F_k \, dJ,
\end{equation}
where we took into account the relation $E_k = 2 \pi k \, \omega_k \, n_k$,
and we took into account the cutoff at $J= J_{Ph}$ related to the fact that for
$J> J_{Ph}$ the nonlinearity is stronger than the linear terms which means
severe damping via wave breaking (i.e. the gravity force is not able to keep
the fluid particles attached to the surface).
This leads to the following estimate of the relationship between the $J-$ and the $k-$ fluxes,
\begin{equation}
F_k \sim {\partial_k \epsilon_k \over  2 \pi k \, \omega_k \, J_{Ph}},
\end{equation}
Thus, the intermittent tail of the PDF (\ref{1/J})   becomes
\begin{equation}
\mathcal{P}_k(J) \approx {F_k \over J \beta_k} \sim
{\partial_k \epsilon_k \over  2 \pi k \, \omega_k \, \beta J \, J_{Ph}}
\sim {n_k \over J \, J_{Ph}}.
 \label{1/J*}
\end{equation}
Here, we used the fact that in the kinetic equation
\begin{equation}
\dot n_k = \alpha_k - \beta_k n_k
 \label{ke}
\end{equation}
the two terms on the RHS are of the same order.

The PDF tail (\ref{1/J*}) gives the following contribution to
 the moments of the Fourier amplitudes
\begin{equation}
M^{(p),tail}_k = \int_0^{J{Ph}} J^p \, \mathcal{P}_k(J) \, dJ
\sim {n_k \over p } \, J_{Ph}^{(p-1)}.
 \label{Mp}
\end{equation}
For example, for the ZF states $n_k = n_{ZF}$, we have
\begin{equation}
M^{(p),tail}_k
\sim {1 \over p } \, \epsilon^{1/3}g^{(p-1)/2} \, k^{1/2 - 9p/2} .
 \label{Mp'}
\end{equation}

On the other hand, the PDFs core has a Rayleigh shape, $$ \mathcal{P}_k(J)
\approx {1 \over n_k} \, e^{ -J/n_k}, $$ which corresponds to Gaussian
statistics of the wave field. The core part gives the following contribution to
the moments, \begin{equation} M^{(p),core}_k = p! n_k^p.
 \label{Mp_core}
\end{equation} For the ratio of the tail and core contributions we have \begin{equation} M^{(p),tail}_k/M^{(p),core}_k \sim {1 \over p p!} \, \left(
J_{Ph}/n_k \right)^{(p-1)}.
 \label{Mp_ratio}
\end{equation} In particular, for the ZF state \begin{equation} M^{(p),tail}_k/M^{(p),core}_k \sim {1 \over p p!} \, \left({k \over k_*}
\right)^{(1-p)/2},
 \label{Mp_ratioZF}
\end{equation} where $k_* = g \epsilon^{-2/3}$.

Thus we can see that for a fixed $p>1$ and at a fixed $k$, the PDF tail will dominate in the moments as $\epsilon \to 0$. On the other hand, at fixed
$\epsilon$ and $k$, the core will dominate when $p \to \infty$.

\section{Experimental setup.}

\begin{figure}
\includegraphics[width=10cm]{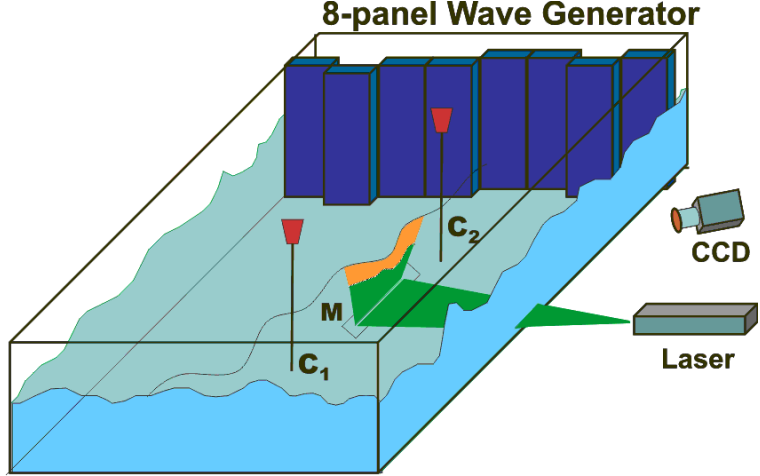}
\caption{\label{fig:Fig1} The experimental setup. M - the first surface mirror, C$_1$ and C$_2$ - the capacitance wire probes, CCD - the digital camera.}
\end{figure}

The  experiments
were conducted in a rectangular tank with dimensions 12 x 6 x 1.5 meters filled
with water up to the depth of 0.9 meters, see Fig. \ref{fig:Fig1}. The gravity waves were
excited by a piston-type wavemaker. The wavemaker consists of 8 vertical
paddles of width 0.75 m covering the full span of one short side of the tank.
An amplitude, frequency and phase can be set for each panel independently
enabling to control directional distribution of the generated waves. A motion
controller is used to program parameters of the generated wave field by
specifying its amplitude and a number of wavevectors (given by a set of
frequencies and directions). In the experiments described here the wavemaker
generated a superposition of two waves  of equal amplitude with frequencies
$f_1$=0.993Hz and $f_2$=1.14Hz (the wavelength is about . The wavevector $k_1$
was perpendicular to the plane of the wavemaker and $k_2$ was at the angle
$7^\circ$ to $k_1$. It is assumed that energy dissipation is low and the waves
undergo multiple reflections from the flume walls, interact to each other and
form a chaotic wave field homogeneous in the central area of the flume. The
main control parameter was an oscillation amplitude of the wavemaker, by
varying it we study the dependence of the spectrum and PDFs on the average wave
intensity.

Two capacitance wire probes were used to measure the wave elevation as a
function of time $\eta(t)$ in two fixed points in the central part of the flume
(as it shown in Figure \ref{fig:Fig1}). The distance between the probes was 2m. Signals from
the probes were amplified and digitized
by a 16-bit analog-to-digital converter (NI6035) controlled by the LabView and
stored in a PC. Typical signal acquisition parameters were as follows: the
bandwidth - 32 Hz
and the recording time - 2000 seconds. The wire probes were calibrated before
the measurements in the same tank with a stationary water surface.

In addition to measurements of the time dependance, in the present work we
introduced a new technique, similar to \cite{Mukto} which allows us to measure
the dependence of the surface elevation on the space coordinate along a line.
For this we used a vertical cross section image of the air-water interface. The
upper layer of water was colored by a fluorescent dye Rhodamine 6G. The
water-air interface area was illuminated from below by a narrow light sheet
from a pulsed Yag laser (power 120 mW, wavelength 532 nm), see Figure
\ref{fig:Fig1}. The images were captured by a 1.3 Mpixel digital camera
(Basler, A622f) synchronously with laser pulses at the sampling frequency 8 Hz.
The image size is 900 x 1200 mm with the resolution 0.93 mm/pixel and 0.90
mm/pixel in the vertical and the horizontal directions respectively. Typically,
we collected five sets of images, each set consisting of 240 frames. The time
interval between the sets was 5 minutes. The data from the capacitance probes
were acquired continuously and in parallel with the images during this time.
The measurements were done at fixed excitation parameters. The measurement
procedure included setting the amplitude of the wavemaker oscillations, waiting
for a transient time interval, 20-30 min, and recording the signals during 35
min.

The data were processed using the Matlab. The wire probe data were filtered by
a band-pass filter within 0.01-20Hz frequency band.
%
The image sets were processed using standard binarization and the boundary
detection procedures from the Image Processing Toolbox. Detected air-water
boundaries were stored as a set of boundary arrays $\eta(x)$ for a following
statistical analysis. The images where the boundary was not a single valued
function of $x$ or when it had significant jumps ($|\delta\eta(x)|/\delta
x|>4$) were deleted. A proportion of such images was less than 3\%. To
calculate spectra from wire probes we used the Welch algorithm with the Hamming
window and the averaging performed over 1000 spectral estimates for each signal
record. The $k-$spectra were calculated for each array of boundaries (one array
from each image) and then averaged over a set from up to 1200 images for each
stationary wave field.

As a characteristics of the averaged wave amplitude we used a nonlinearity
parameter which is defined as the mean slope of the wave at the energy
containing scale, $\gamma = k_m A,$ where  $k_m$ is the wavenumber
corresponding to the maximum of the energy spectrum $|\eta_\om|^2$. In all our
experiments $k_m$ was approximately the same and located in the forcing range,
$k_m \approx 5.7 $m$^{-1}$ that corresponds to the wavelength $\lambda \approx
1.1$ m. In this experiment the range of the nonlinearity parameter was
$0.1<\gamma<0.25$.

\section{Results.}

\subsection{Spectra.}

\begin{figure}
\includegraphics[width=\linewidth]{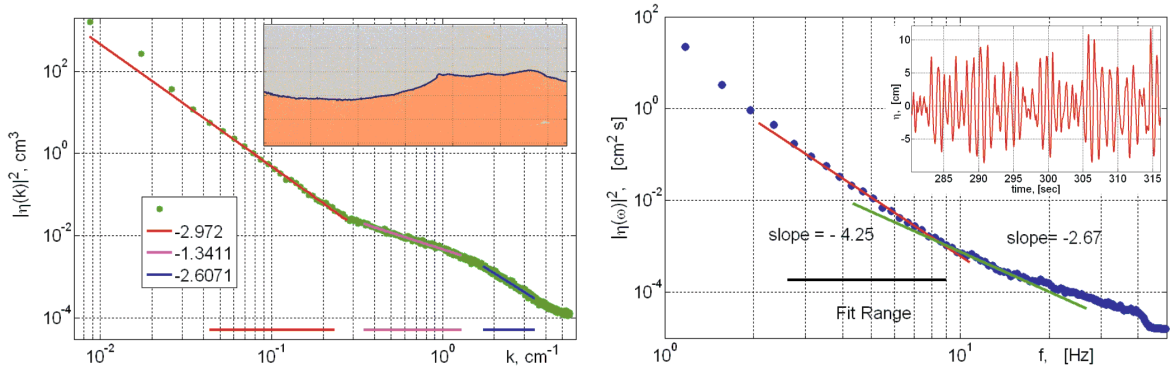}
\caption{\label{fig:kwspectra} Left plot: The spectrum in $k$-domain. The inset shows an image of air-water interface. Right plot: The spectrum in $\omega$-domain. The inset shows a correspondent function $\eta(t)$. Both spectra were measured at the wave nonlinearity $\gamma$=0.2.}
\end{figure}

Typical energy spectra in the $\om$ and $k$ domains are shown in Figures \ref{fig:kwspectra}.
For the $\om$-spectra we usually have about one decade of the fitting range
which, according to the dispersion relation, should correspond to the two
decades for the $k$-spectra. In reality the $k$-spectra have shorter scaling
range which is limited, on the low $k$-side, by the width of laser sheet,1.2m,
and on the high $k$-end by insufficient vertical resolution of the images and
limited statistics. In addition, the scaling ranges getting narrower for the
flume runs with weaker forcing.

\begin{figure}
\includegraphics[width=\linewidth]{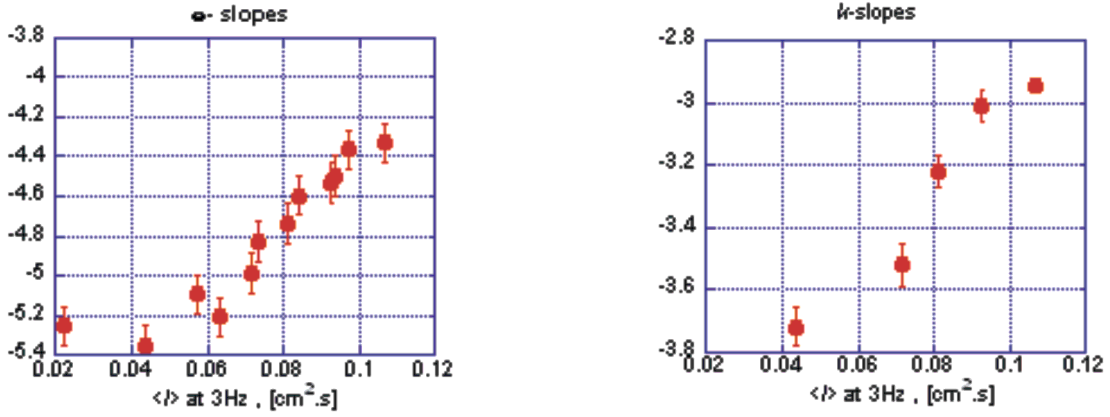}
\caption{\label{fig:w_and_k} Slopes of the $k$- and $\om$-spectra as functions of the wave intensity, $I=|\eta_\omega|^2$, measured at the frequency 3Hz in $\om$-spectrum.}
\end{figure}

The slopes of the energy spectra in the $\om$ and $k$-domains as functions of
the wave field intensity are shown in Fig. \ref{fig:w_and_k}.
We see that for both $\om$ and $k$ spectra
 the slopes are steeper for the weaker wave fields with respect to the
stronger ones. One can see that at low wave intensities the data scatter and
uncertainty are much greater than for stronger wave turbulence in agreement
with the previous measurements of the $\om$ spectra in \cite{DLN07}.

\begin{figure}
\includegraphics[width=8cm]{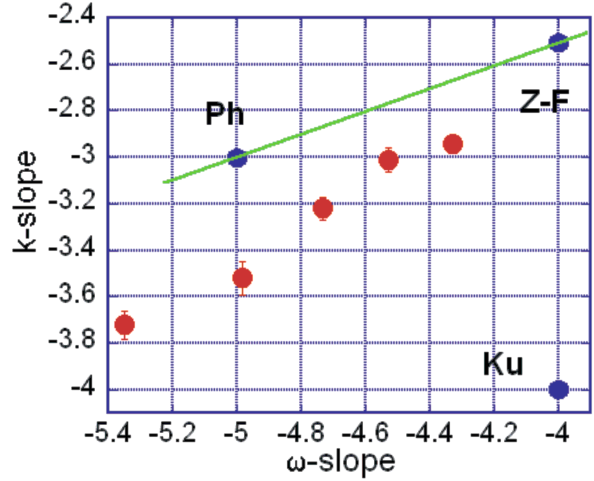}
\caption{\label{fig:kw_slopes} The $k$-slope vs $\om$-slope. The linear
dispersion relation, $\om = \sqrt{gk}$, is shown by the solid line. Ph, ZF and
Ku mark the points corresponded to Phillips, Zakharov-Filonenko and Kuznetsov
predictions respectively.}
\end{figure}

In Figure \ref{fig:kw_slopes} we show of graph of the $k$-slope versus the
$\om$-slope for the energy spectra measured in the same experiments with the
laser sheet and the capacitance wire techniques respectively. The linear
dispersion slope is shown by the solid curve. As we see, the experimental data
deviate significantly from the linear dispersion curve, which indicates that,
at least in the fitting ranges of scales, the nonlinearity is not weak.

We also put the points corresponding to the theoretical predictions: Ph, ZF and
Ku spectra. We see that both Ph and Ku points are rather far from the
experimental data, whereas ZF point is more in agreement with the experiment.
Namely, this plot suggests that if one would perform experiments at even higher
amplitudes then it is quite likely that the experimental data would have
crossed the ZF point. This result is rather surprising because one would
naturally expect the ZF theory to work better for weaker rather than for
stronger waves. However, as we mentioned before, the finite flume size effects
are more important for weaker waves, which is the most likely explanation why
there is a significant deviation from the ZF spectrum at smaller wave
amplitudes. Note that the simultaneous measurement of the $k$ and $\om$-spectra
allows one to resolve the uncertainty of the previous results reporting on the
the $\om$-spectra only. Namely, we are now able to differentiate between the ZF
and Ku states which have undistinguishable $\om$-slopes but different
$k$-slopes. The result is that ZF spectrum is more in agreement with the
experimental data than the Ku spectrum. However, as we will see, coherent
singular structures of the type discussed by the Ku theory do seem to leave
their imprints on the scalings of the high order structure functions.

\begin{figure}
\includegraphics[width=10cm]{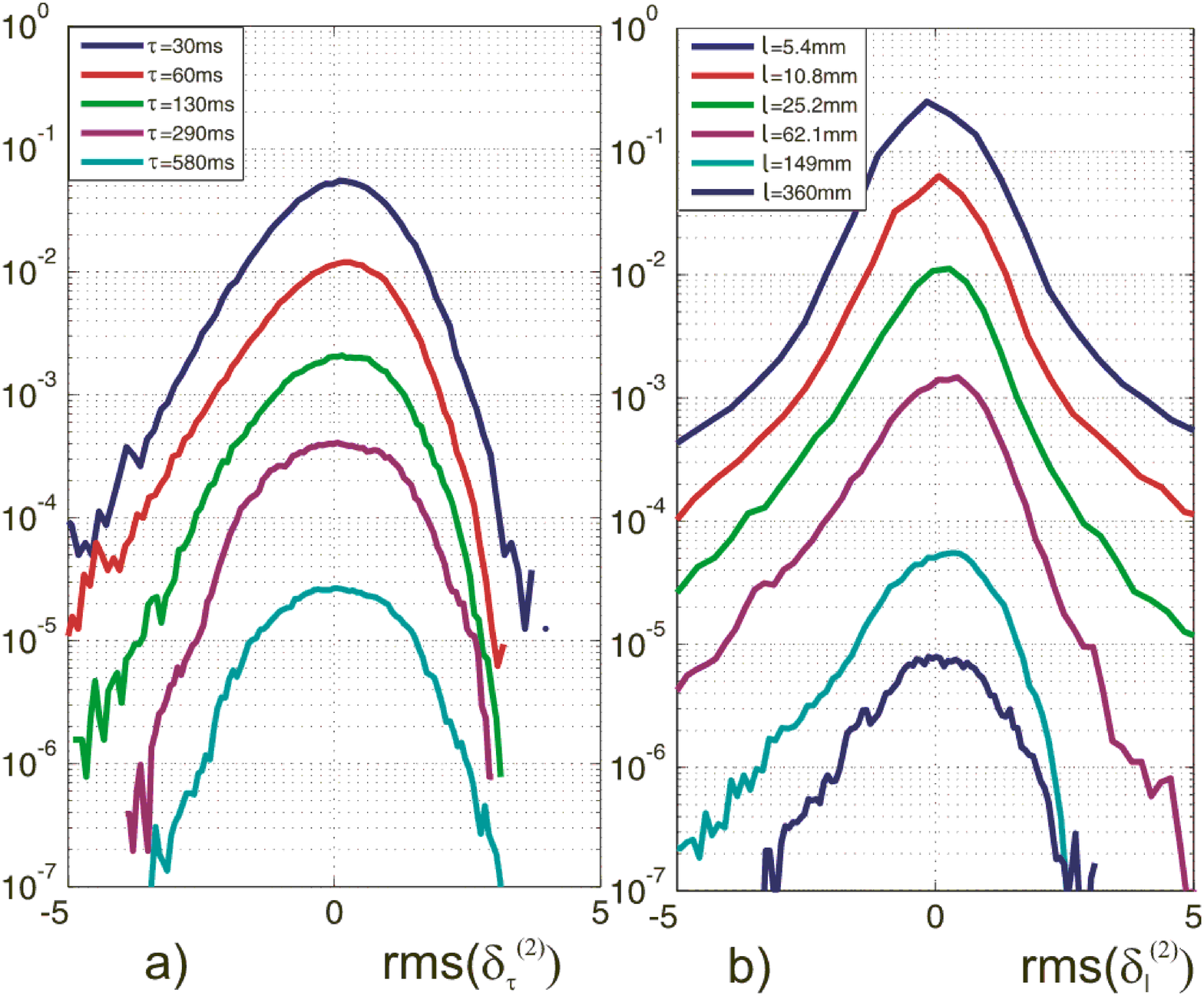}
\caption{\label{fig:pdfs} PDFs of second-order differences: a) in the $t$-domain, b) in the $x$-domain.}
\end{figure}

\subsection{PDFs of the height increments.}

To present results on the PDFs and the SFs we select the experimental run with
spectra
$E_k \sim k^{-3.02}$ and $ E_\om \sim \om^{-4.53}$.
Because for each of these spectra both $k$ and $\om$ slopes are steeper than -3
but shallower than -5, we choose to work with the second-order increments,
$j=2$. Experimental PDFs of the height increments in space and time are shown
in Figs. \ref{fig:pdfs}a and \ref{fig:pdfs}b respectively. For the space
increments, one can see clearly deviations from Gaussianity at the (fat) PDF
tails  which are related to intermittency and indicate  presence of the
coherent structures.
For the time increments, the deviations from Gaussianity is much less
pronounced, which could be due to the slow propagation speed of the coherent
structures leading to their more infrequent occurrence in the $t$-domain in
comparison with the $x$-domain. Both the $t$- and the $x$-domain PDF's are
asymmetric (with the negative increments dominant) which results from breaks
occurring at wave crests rather than troughs.

\subsection{Structure functions.}

\begin{figure}
\includegraphics[width=10cm]{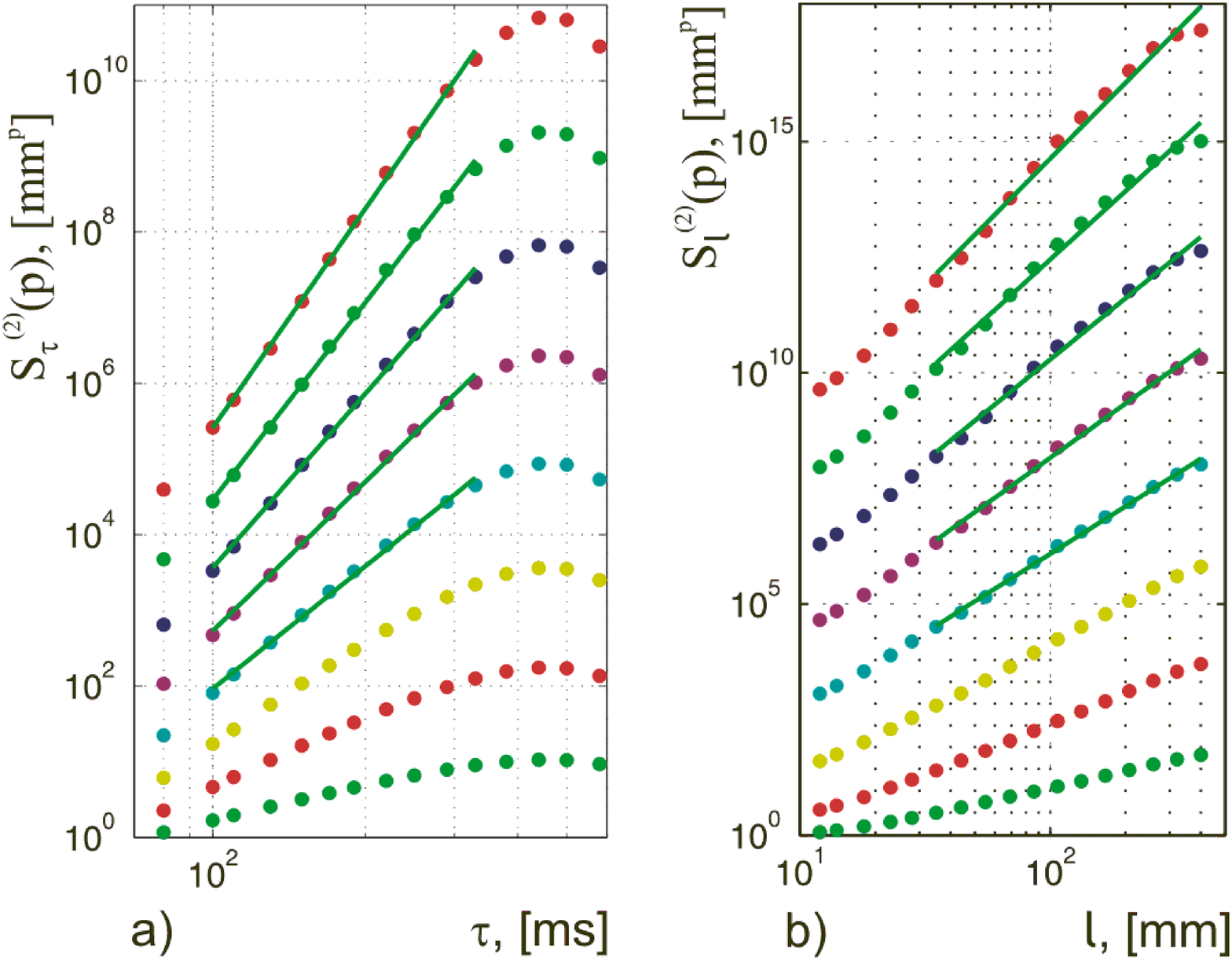}
\caption{\label{fig:moms} The elevation structure functions (moments) from the 1st to 8th order (p=1, ... ,8): a) in the $t$-domain, $S^{(2)}_\tau (p)$, and b) in the $x$-domain, $S^{(2)}_l (p)$, as the functions of $\tau$ and $l$ respectively.}
\end{figure}

\begin{figure}
\includegraphics[width=10cm]{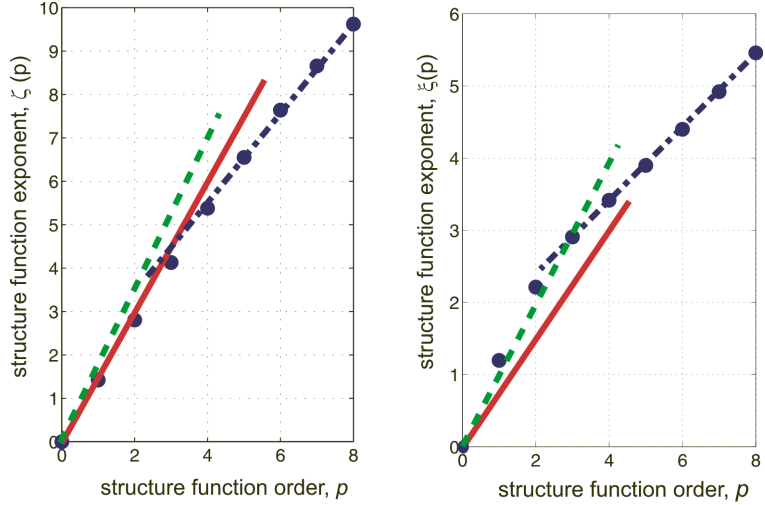}
\caption{\label{fig:momexp} Structure functions scaling exponents a) $\xi(p)$
in the $k$-domain and b)$\zeta(p)$ in the $t$-domain. Red-solid line represents ZF scaling, green-dash line - the scaling obtained in the experiment, blue-dash-dot line - the scaling for coherent structures.}
\end{figure}

In our data on the SFs, both $S^{(2)}_\tau(p)$ as a function of $\tau$ and
$S^{(2)}_l(p)$ as a function of $l$ exhibit clear power-law scalings in the
range of scales corresponding to the gravity waves for  all $p$ at least up to
8 (see Fig.~\ref{fig:moms}).
The SF exponents for the time and space domains are shown in
Figs.~\ref{fig:momexp}a and \ref{fig:momexp}b respectively. Straight lines on
these graphs represent the ZF scaling (solid line, red online),
scaling of waves with the spectrum as measured in the experiment (dash line,
green online) and the fit of the high-$p$ behavior with a scaling corresponding
to singular coherent structures
(dash-dot line, blue online).
For the time domain, the scaling at low $p$ is close to the  ZF scaling, this
is surprisingly more consistent than with the scaling calculated from the
actual measured spectrum. For an infinite scaling range the $p=2$ point must,
of course, lie exactly on the value corresponding to the spectrum, $(\nu -1)$,
irrespective of the presence or absence of the phase  correlations. Thus we
attribute the observed discrepancy to the finiteness of the scaling range.
Furthermore, the fit of the high $p$ dependence  indicates the presence of
singular coherent structures with $D=1, a=1.05$ that  is very close to the Ku's
$D=1, a=1$.

For the space-domain, at low $p$ there is an agreement with the scaling of the
random phased waves having the actual measured spectrum, and less agreement
with the random phased waves having the ZF spectrum. This is not surprising
since the scaling range in $k$ is greater than in $\omega$ and therefore there
is a better agreement between the spectrum and the SF exponent for $p=2$. More
importantly  we see again the dominance of the random phased waves in the
low-order SFs, and the dominance of coherent breaks in the high-order SFs. The
fit at high $p$'s gives for the dimension and the singularity parameter of the
breaks $ D = 1.3$ and $ a = 1/2$ respectively. We see that
the breaks appear to be more singular and "spiky" than the Ku-type breaks
(a = 1).
Visually, we observed numerous occurrences of these kinds of spiky wave breaks,
which are not propagating (or propagating very slowly)
and producing vertical splashes. These kinds of structures should be probable
in isotropic wave fields due to the collision of counter-propagating waves,
which in our flume appear due to multiple wave reflections from the walls. The
slow propagation speed of such breaks means  their seldom crossing through the
capacitance probe  even if there is a large number of them in the
$x$-domain (i.e. more than the Ku-type breaks). This could explain why
 the Ku-breaks show up in the SF
scalings in the $t$-domain, whereas more singular spiky structures are seen in
the $x$-domain.

\subsection{Statistics of the Fourier modes.}

Let us now present results on the statistics of the Fourier modes. Immediately,
we should be cautions about the applying the theory of the Fourier mode
statistics and intermittency cycle described above in the theoretical part of
the paper. Firstly, the theoretical picture was developed for weak waves where
random waves dominate over the coherent structures in most of the inertial
range, whereas the experimental runs selected in the present work corresponded
to larger excitations where the rms amplitude and wavebreaking amplitude are
not different by orders of magnitude. Secondly, like classical WTT, the theory
corresponds to an infinite system, whereas the finite size effects are likely
to be important in our flume. Recall that the finite size effects are strongest
at low amplitudes, and it is impossible to implement weak wave turbulence and
eliminate the finite size effects simultaneously in our flume. Thirdly, our
statistical data is not sufficient for the single Fourier modes and, as a
result, the PDF tails are rather noisy. One can reduce this noise by combining
several adjacent  Fourier modes, but in doing this one has to be careful not to
reduce or eliminate the tail due to such an averaging (i.e. one can only
combine modes which are strongly correlated).

\begin{figure}
\includegraphics[width=10cm]{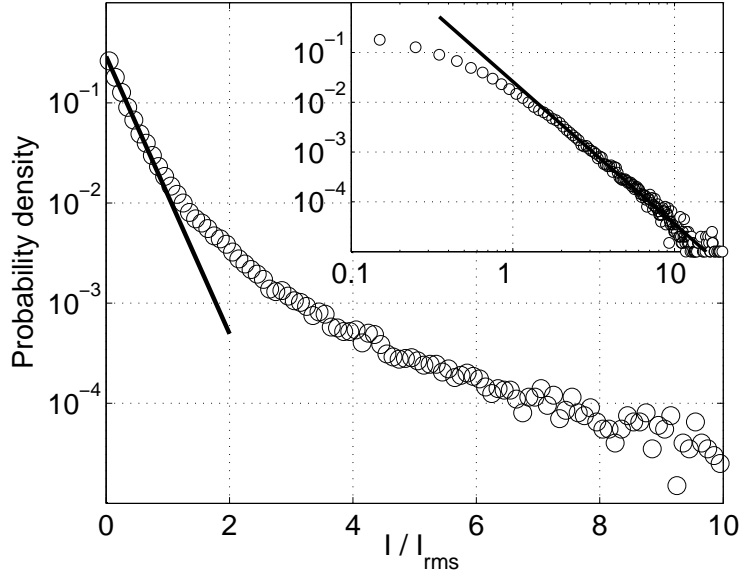}
\caption{\label{fig:w-pdf} Normalized PDF of the spectral intensity
$I=|\eta_w|^2$ band-pass filtered with the frequency window $\pm 1Hz$ centered
at $f=$6 Hz, measured at wave nonlinearity $\gamma=k_0\eta_{rms}\approx$0.16
($k_0$ is the wave vector at the maximum spectral power). The inset shows the
same plot in log-log coordinates.}
\end{figure}

Nonetheless, the results on Fourier space PDFs are quite illuminating and worth
presenting. PDFs of $\omega$-modes, $J_\omega = |\eta_\omega|^2$,
 as measured by the wire probes were first presented
in \cite{DLN07} and we reproduce them in Figure \ref{fig:w-pdf}. Here, the
averaging was done over modes in the frequency range from 5.5 to 6.5 Hz via
bandpass filtering. One can see that the PDF core can be fitted with an
exponential function which corresponds to gaussian statistics. At the tail one
can see a significant deviation from the exponential fit corresponding to
intermittency. As predicted by the theory, the tail follows a power law (with
cutoff), but with a different index, $-3$ rather than $-1$. This deviation in
the power-law index could be due to one of the reasons mentioned in the
beginning of this subsection.

\begin{figure}
\includegraphics[width=\linewidth]{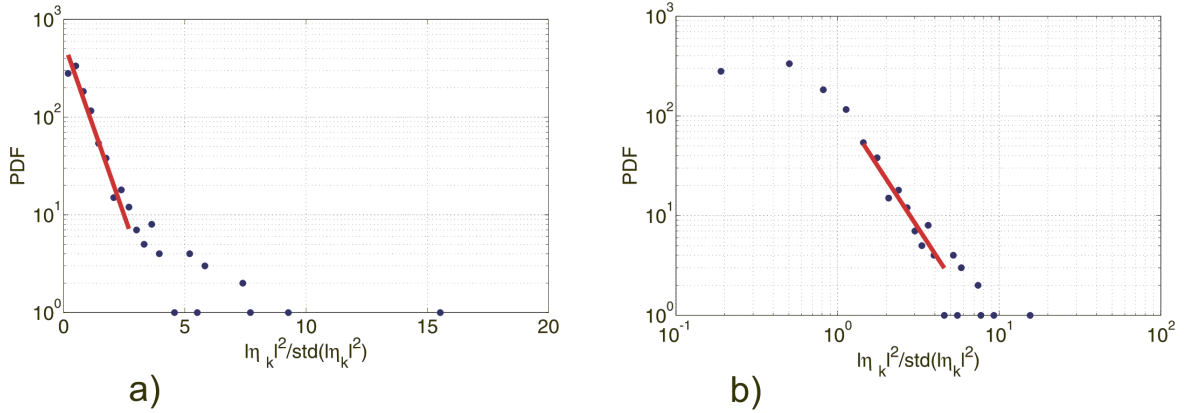}
\caption{\label{fig:k-pdf} The PDF of the $k$-mode centered at $k$=54.3rad/m,
filtered within the window $\pm 10.8$rad/m and normalized by its standard
deviation. a)A semilog plot with the exponential fit $y\propto 10^{-1.6x}$ b) A
loglog plot with the power fit $y\propto x^{-2.5}$.}
\end{figure}

PDFs of $k$-modes, $J_k = |\eta_k|^2$,
 are shown in Figure \ref{fig:k-pdf}.
Here, the averaging was done over 3 adjacent modes. Again, we see a Gaussian
PDF core and a power-law tail, - now with index $-2.5$, which also different
from the theoretical index $-1$. The fact that the PDF tail for the $k$-modes
decays slower than for the $\omega$-modes ($k^{-2.5} $ vs $k^{-3} $) is
consistent with our conclusion (which we have made based in the statistics of
the $x$ and $t$-domain increments) that turbulence shows stronger intermittency
in space than in time.

\section{Summary and discussion.}

We have presented a theory and performed measurements of gravity wave turbulence
statistics in both $x$ and $t$-domains.
This allowed us to differentiate between the states which have the same
frequency spectra but different $k$-spectra, particularly between the ZF and Ku
states.
Our data indicate that the spectral
exponents, both in $\omega$ and $k$, depend on the amplitude of the forcing.
There is a certain evidence favoring ZF theory at larger wave amplitudes,
but the ZF state in not possible to observe in its purity because the two
fundamental limits of WTT, weak nonlinearity and the infinite box, are impossible
to implement simultaneously even in such rather large flume as ours.
None of the existing theories, relying on the presence of either random phased
weakly nonlinear waves or on dominance of coherent wave crests of a particular
type, can fully explain our experimental results. Instead, there is an indication that the
gravity wave field consists of coexisting and interacting random and coherent
wave components. The random waves are captured by the PDF cores and the
low-order SF's, whereas the coherent wave crests leave their imprints on the PDF
tails and on the high-order SF's. The singular wave crests themselves consist of
structures of different shapes: numerous non-propagating spikes/splashes (which
show in the $x$-domain SF's) and propagating Ku-type breaks (seen in the
$t$-domain SF's). We suggested a plausible scenario for
 the dynamics and mutual interactions of these
coexisting random-phased and coherent wave components based
on a turbulence cycle. Namely, the coherent structures appear in the
process of the energy cascade within the random wave component (when
the nonlinearity becomes strong at some scale along the cascade).
The coherent waves partially dissipate their energy due to wave breaking
and partially they return energy to a wide range of longer incoherent waves.
Based on this picture, we made qualitative theoretical predictions for the
scalings of the moments of the Fourier modes. However, our present statistical data
is insufficient for building these moments, and longer experimental runs are
needed in future to accumulate this statistics.

\section{Ackowledgements.}

This work was partially done during the EPSRC sponsored Warwick Turbulence
Symposium programme 2005-2007. The authors are grateful to the Hull
Environmental Research Institute for a partial financial support and B. Murphy
for his help with experiments at the Deep flume.

\end{document}